\def\qddd{\lower2pt\hbox{$\stackrel{\textstyle .\kern-.1em . \kern-.1em .}{q}$}}
\documentclass[12pt]{article}
\input{psfig.sty}
\begin{document}
\begin{titlepage}
\begin{flushright}
hep-th-/9707088 \\
UFIFT-HEP-97-19 \\
NBI-HE-97-25 \\
July, 1997
\end{flushright}
\vspace{.4cm}
\begin{center}
\textbf{The Initial Value Problem For Maximally Non-Local Actions}
\end{center}
\begin{center}
D. L. Bennett$^{\dagger}$ \\
and \\
H. B. Nielsen$^{\#}$
\end{center}
\begin{center}
\textit{Niels Bohr Institute \\ DK-2100 Copenhagen \O \\ DENMARK}
\end{center}
\begin{center}
R. P. Woodard$^*$
\end{center}
\begin{center}
\textit{Department of Physics \\ University of Florida \\ 
Gainesville, FL 32611 USA}
\end{center}
\begin{center}
ABSTRACT
\end{center}
\hspace*{.5cm}
We study the initial value problem for actions which contain non-trivial
functions of integrals of local functions of the dynamical variable. In
contrast to many other non-local actions, the classical solution set of 
these systems is at most discretely enlarged, and may even be restricted,
with respect to that of a local theory. We show that the solutions are 
those of a local theory whose (spacetime constant) parameters vary with 
the initial value data according to algebraic equations. The various roots 
of these algebraic equations can be plausibly interpreted in quantum 
mechanics as different components of a multi-component wave function. It is 
also possible that the consistency of these algebraic equations imposes 
constraints upon the initial value data which appear miraculous from the 
context of a local theory.
\begin{flushleft}
\end{flushleft}
\vspace{.4cm}
\begin{flushleft}
$^{\dagger}$ e-mail: bennett@nbivms.nbi.dk \\
$^{\#}$ e-mail: hbech@nbivms.nbi.dk \\
$^*$ e-mail: woodard@phys.ufl.edu
\end{flushleft}
\end{titlepage}

Most physicists probably recall wondering, at their first exposure to 
Lagrangians, why they are usually assumed to depend only upon the zeroth and 
first time derivatives of the dynamical variable. The answer is very simple: 
allowing higher time derivatives almost always leads to instabilities. This has
been known since Ostrogradski's canonical formulation of such systems in the 
middle of the 19th century \cite{oy}.

To understand the problem, consider the dynamics of a point particle in one 
dimension whose position is $q(t)$ and whose Lagrangian $L(q,\dot{q},\ddot{q})$
includes second time derivatives. We assume only that the second derivatives
cannot be removed by partial integration. This condition is known as {\it
non-degeneracy} and amounts to the invertibility of the equation:
\begin{equation}
P_2 = {\partial L(q,\dot{q},\ddot{q}) \over \partial \ddot{q}} \; ,
\end{equation}
to solve for $\ddot{q}(q,\dot{q},P_2)$. Under the assumption of non-degeneracy 
the Euler-Lagrange equations:
\begin{equation}
{\partial L \over \partial q} - {d \over d t} {\partial L \over \partial 
\dot{q}} + \left({d \over d t}\right)^2 {\partial L \over \partial \ddot{q}}
= 0 \; , \label{eq:EL}
\end{equation}
define time evolution by determining the fourth derivative of $q(t)$ in terms
of $q$, $\dot{q}$, $\ddot{q}$ and $\qddd$. One must therefore supply twice as 
much initial value data as in the usual case, and this entails a new set of
canonically conjugate coordinates in the Hamiltonian formulation. In 
Ostrogradski's construction the canonical coordinates are:
\begin{equation}
Q_1 \equiv q \hspace{2cm} Q_2 \equiv \dot{q} \; ,
\end{equation}
and they are respectively conjugate to the following momenta:
\begin{equation}
P_1 \equiv {\partial L(q,\dot{q},\ddot{q}) \over \partial \dot{q}} - {d \over
dt} {\partial L(q,\dot{q},\ddot{q}) \over \partial \ddot{q}} \hspace{2cm}
P_2 \equiv {\partial L(q,\dot{q},\ddot{q}) \over \partial \ddot{q}} \; .
\end{equation}
Note the equation for $P_2$ can be inverted to solve for $\ddot{q}$ in terms of
just $Q_1$, $Q_2$ and $P_2$; $P_1$ is only needed to express $\qddd$.

Ostrogradski's Hamiltonian is:
\begin{eqnarray}
H & = & \sum_{i} P_i \dot{Q}_i - L \\
& = & P_1 Q_2 + P_2 \ddot{q}(Q_1,Q_2,P_2) - L(Q_1,Q_2,\ddot{q}(Q_1,Q_2,P_2)) 
\; , \label{eq:Ostro}
\end{eqnarray}
and his canonical equations for time evolution are the obvious ones suggested 
by the notation:
\begin{equation}
\dot{Q}_i = {\partial H \over \partial P_i} \hspace{2cm} \dot{P}_i = -{\partial
H \over \partial Q_i} \; .
\end{equation}
It is straightforward to verify that the evolution equations for $Q_1$, $Q_2$ 
and $P_2$ simply reproduce the definitions of $Q_2$, $P_2$ and $P_1$, 
respectively. The canonical expression of the Euler-Lagrange equation 
(\ref{eq:EL}) is the evolution equation for $P_1$. So Ostrogradski's 
Hamiltonian generates time evolution; it is also conserved when the Lagrangian 
is free of explicit time dependence. 

The problem with higher derivatives is apparent from expression 
(\ref{eq:Ostro}): {\it the Hamiltonian is linear in} $P_1$. Such a Hamiltonian
can never be bounded below; in fact there is not even any barrier to the 
system's decay. In a conventionally interacting system this means that one can
excite positive energy degrees of freedom while conserving energy by exciting
negative energy degrees of freedom. Since there are typically many more ways of
exciting a degree of freedom than not, the system will tend to migrate to very
highly excited states which have little in common with the physical reality we
perceive.

Note the generality of the higher derivative instability. It does not depend 
upon any approximation scheme, nor any feature of the Lagrangian except 
non-degeneracy. Nor is quantization liable to prevent it because the 
instability is not confined to a small region of phase space. Note also the
relation to the space of classical solutions: a non-degenerate higher 
derivative doubles the number of continuum degrees of freedom and at least half
of the new degrees of freedom access negative energy.

Physicists are an inventive lot and such a bald no--go theorem provokes them to
envisage tortuous evasions. It is impossible to prove a negative, so we will 
not assert that there is no way out for non-degenerate higher derivatives, but
we do urge a little common sense. Ostrogradski's theorem should not seem 
surprising. It explains why every single system we have so far observed seems 
to be described, on the fundamental level, by a Lagrangian containing no higher 
than first time derivatives. The bizarre and incredible thing would be if this 
fact was simply an accident.

The relevance of Ostrogradski's theorem to non-local actions is that the 
instability grows worse as more higher derivatives are added. For Lagrangians
which depend non-degenerately upon the $N$-th time derivative the associated
Hamiltonian is linear in all but possibly one of the $N$ canonical momenta
\cite{oy}. This is disastrous for non-local Lagrangians which can be 
represented as limits of increasingly higher derivative Lagrangians. Such a 
representation is valid when the non-locality enters through entire functions 
of the derivative operator which cannot be subsumed into a field redefinition. 
A familiar example of such a system is string field theory \cite{ew}. On the 
other hand, certain forms of non-locality are innocuous. For example, one 
generally obtains a non-local Lagrangian by integrating out one or more of the 
fundamental dynamical variables. These Lagrangians contain poles of the 
derivative operator, so it is not valid to consider them as limits of higher 
derivative Lagrangians. And since it is of course valid to solve them in the 
original, local form, there is no extension of the space of classical 
solutions.

A local action is defined as the integral of a function of the dynamical
variable and some finite number of its derivatives. What we will call a {\it 
maximally non-local action} involves non-trivial functions of such terms.
As an example, consider the following generalization of the one-dimensional 
harmonic oscillator:
\begin{equation}
S[q] = \lim_{T \rightarrow \infty} \left\{\int_{-T}^T dt \; \frac12 m 
\dot{q}^2 - \frac14 m \omega_0^2 \ell_0^2 T \left({1 \over T} \int_{-T}^T dt
\; {q^2 \over \ell_0^2}\right)^2 \right\} \; , \label{eq:harmo} 
\end{equation}
where $\omega_0$ and $\ell_0$ are constants with the respective dimensions of
frequency and length. An important fact about maximally non-local actions is
that their equations of motion are local except for coupling ``constants'' 
which are functions of integrals of local functions of the dynamical variable. 
For the example just presented one finds:
\begin{equation}
{\delta S[q] \over \delta q(t)} = - m \ddot{q}(t) - m \omega^2[q] q(t) = 0 \; ,
\label{eq:Euler}
\end{equation}
where the oscillator's frequency is:
\begin{equation}
w^2[q] \equiv \lim_{T \rightarrow \infty} {\omega_0^2 \over T \ell_0^2} \int_{
-T}^T dt \; q^2(t) \; . \label{eq:omega}
\end{equation}
It has been argued that quantum field theoretic versions of maximally non-local
actions might explain the apparent fine tuning of certain coupling constants 
\cite{bfn}. Of course any such mechanism would be pointless if it inescapably 
entailed the Ostrogradskian instabilities. The aim of this paper is to show 
that it does not.

It is simplest to begin by solving the maximally non-local harmonic oscillator
(\ref{eq:harmo}). From its definition (\ref{eq:omega}), $\omega^2[q]$ must be 
positive semi-definite, so the general solution for fixed $\omega$ is:
\begin{equation}
q(t) = q_0 \cos(\omega t) + {\dot{q}_0 \over \omega} \sin(\omega t) \; ,
\label{eq:harmonic}
\end{equation}
where $q_0$ and $\dot{q}_0$ are the initial value data. Substituting this into
the definition of $\omega^2$ gives:
\begin{equation}
\omega^2 = {\omega_0^2 \over \ell_0^2} \left(q_0^2 + {\dot{q}_0^2 \over 
\omega^2}\right) \; .
\end{equation}
This equation has two solutions but only the positive root is consistent with
the explicit positive semi-definiteness of ({\ref{eq:omega}). We can therefore
write:
\begin{equation}
\omega^2 = {\omega_0 \over 2 \ell_0^2} \left(\omega_0 q_0^2 + \sqrt{\omega_0^2 
q_0^4 +  4 \ell_0^2 \dot{q}_0^2 }\right) \; .
\end{equation}

Since this theory is time translation invariant, it makes no difference if we 
replace $q_0$ and $\dot{q}_0$ in $\omega^2[q]$ by $q(t)$ and $\dot{q}(t)$,
respectively. The vanishing of $d\omega^2/dt$ follows from the equation of
motion (\ref{eq:Euler}). The conserved energy associated with time translation
invariance is:
\begin{equation}
E = \frac12 m \dot{q}^2 + \frac12 m \omega^2 q^2 \; . \label{eq:energy}
\end{equation}
To make it generate time evolution we define the Poisson bracket as follows:
\begin{eqnarray}
\{q,\dot{q}\} & = & \dot{q} \div {\partial E \over \partial \dot{q}} \\
& = & {1 \over m } {\sqrt{\omega_0^2 q^4 + 4 \ell_0^2 \dot{q}^2} \over 
\sqrt{\omega_0^2 q^4 + 4 \ell_0^2 \dot{q}^2} + \omega_0 q^2} \; .
\end{eqnarray}
This makes the Poisson bracket of $q$ with $E$ give $\dot{q}$. The other 
evolution equation:
\begin{equation}
\ddot{q} = \{\dot{q},E\} = - {\partial E \over \partial q} \times 
\{q,\dot{q}\} \; ,
\end{equation}
follows by taking the time derivative and using energy conservation. It can 
also be checked explicitly.

Since the energy (\ref{eq:energy}) of this system is bounded below we have 
shown by explicit example that the Ostragradskian instabilities are not
inevitable for maximally non-local actions. For the system considered there
was not even any enlargement of the solution set of its local cognate --- 
there is a unique solution for every choice of $q_0$ and $\dot{q}_0$. This
feature is not generic; more complicated models can show a discrete enlargement
of the solution set. Consider, for example, the following maximally non-local
action:
\begin{equation}
S[q] = \lim_{T \rightarrow \infty} \left\{\int_{-T}^T dt \; \left[\frac12 m 
\dot{q}^2 - \frac14 m \omega^2_0 q^2\right] - \frac18 m \omega^2_0 \ell_0^2 T
\sin\left[ {2 \over T} \int_{-T}^T dt {q^2 \over \ell_0^2}\right] \right\} 
\; .
\end{equation}
The field equations are again those of a harmonic oscillator but with a 
slightly different form for the frequency-squared:
\begin{eqnarray}
{\delta S[q] \over \delta q(t)} = - m \left\{ \ddot{q}(t) + \omega^2[q] q(t)
\right\} \; , \\
\omega^2[q] \equiv \omega_0^2 \cos^2\left[ \lim_{T \rightarrow \infty} {1 
\over T} \int_{-T}^T dt {q^2 \over \ell_0^2}\right] \; .
\end{eqnarray}
For fixed $\omega$ the general initial value solution is still
(\ref{eq:harmonic}) so $\omega^2$ is any positive root of the following 
transcendental equation:
\begin{equation}
\omega^2 = \omega_0^2 \cos^2\left[{q_0^2 \over \ell_0^2} + {\dot{q}_0^2 \over
\omega^2 \ell_0^2}\right] \; .
\end{equation}
For $\dot{q}_0 \neq 0$ there are a countable infinity of solutions. No closed
form can be obtained, but for large integers $N$ they are approximately:
\begin{equation}
\omega_N^2 \approx {2 \over 2 N + 1} {\dot{q}_0^2 \over \pi \ell_0^2} \; .
\end{equation}
Of course the energy is still (\ref{eq:energy}) --- with the new meaning of
$\omega^2$ --- and it is positive semi-definite for each root.

Another curious feature of maximally non-local actions is the possibility 
for {\it restrictions} on the initial value data. It is this property which
might offer an explanation for otherwise miraculous fine tunings \cite{bfn}.
To understand it, consider the following maximally non-local harmonic 
oscillator:
\begin{equation}
S[q] = \lim_{T\rightarrow \infty} \left\{ -\frac12 m \omega_0^2 \ell_0^2
\exp\left[- {1 \over T} \int_{-T}^T dt {\dot{q}^2 \over \omega_0^2 \ell_0^2}
\right] - \frac12 m \omega_0^2 q^2\right\} \; .
\end{equation}
The equation of motion is still that of a harmonic oscillator but with its
frequency-squared defined thus:
\begin{equation}
\omega^2[q] \equiv \omega_0^2 \exp\left[\lim_{T \rightarrow \infty} {1 
\over T} \int_{-T}^T dt {\dot{q}^2 \over \omega_0^2 \ell_0^2} \right] \; .
\end{equation}
One determines $\omega^2$ as a function of the initial value data from the
following transcendental equation:
\begin{equation}
\omega^2 = \omega_0^2 \exp\left[{\omega^2 q_0^2 \over \omega_0^2 \ell_0^2} + 
{\dot{q}_0^2 \over \omega_0^2 \ell_0^2}\right] \; .
\end{equation}
When the ratio $q_0/\ell_0$ is much less than one the exponential grows slowly
enough to intersect the quadratic, and there are two solutions. However, when
$q_0/\ell_0$ is larger than one there is no solution.

In fact the analysis we have just gone through applies as well to field 
theory. The field equations of maximally non-local actions are related to 
local ones whose couplings, $\lambda_1, \lambda_2, \dots$, are really 
non-dynamical constants. Suppose that the local cognate of a maximally 
non-local action has the following general initial value solution:
\begin{equation}
\phi(t,{\vec x}) = \Phi[\phi_0,\dot{\phi}_0](t,{\vec x},\lambda_1,\lambda_2,
\dots) \; . \label{eq:initial}
\end{equation}
Since its maximally non-local cousin has the same field equation (by 
definition) its solutions must be the same, except that the couplings, 
$\lambda_1, \lambda_2, \dots$, are spacetime constant integrals of the 
dynamical variable:
\begin{equation}
\lambda_i = \Lambda_i[\phi] \; . \label{eq:integral}
\end{equation}
Substituting the general initial value solution (\ref{eq:initial}) into these
integrals gives algebraic equations for the couplings:
\begin{equation}
\lambda_i = \Lambda_i\left[\Phi[\phi_0,\dot{\phi}_0](t,{\vec x},\lambda_1,
\lambda_2,\dots)\right] \; .
\end{equation}
There will in general be more than one root, the choice of which represents 
discrete degrees of freedom not present in the local theory. It is also 
possible that some or all of the roots may disappear unless the initial value
data lie within certain regions.

If its local cognate has a conserved energy the same functional will be 
conserved for a maximally non-local action. One will also be able to impose
Poisson brackets to make it generate time evolution. Whether or not this energy
is bounded below will depend upon what the various solutions do to the energy
functional of the local cognate. But our explicit example of the non-local 
oscillator shows that the Hamiltonian can be bounded below, so there is no 
generic instability of the Ostrogradskian type.

We close by proposing a physical interpretation for the discrete degrees of 
freedom associated with multiple roots of the parameter equations. We do not 
feel one is entitled to select a particular root and ignore the others. 
Instead, we believe that in quantum mechanics a natural interpretation for the 
various roots is as different components of a multi-component wave function. 
One could then treat the various roots the way one works with spin or internal 
quantum numbers such as iso-spin. Note that the dynamics of each component 
would be the same except for different couplings. It is tempting to speculate 
that such a formalism might be used to unify the mysterious generational 
structures which appear in elementary particle physics.

\begin{center}
ACKNOWLEDGEMENTS
\end{center}

This work was partially supported by INTAS Grant 93-3316, EF Contract SCI 0340
(TSTS) and by DOE contract 86-ER40272.

\end{document}